\documentclass[english,aps,prl,twocolumn,showpacs,preprintnumbers,amsmath,amssymb]{revtex4-1}
\pdfoutput=1
\usepackage{graphicx}
\usepackage{bm}
\usepackage{times}
\usepackage{slashed}
\usepackage{color}
\usepackage{epsf}
\usepackage{dcolumn}
\usepackage{pstricks}
\usepackage{color}

\def\ls{\mathrel{\lower4pt\vbox{\lineskip=0pt\baselineskip=0pt
           \hbox{$<$}\hbox{$\sim$}}}}
\def\gs{\mathrel{\lower4pt\vbox{\lineskip=0pt\baselineskip=0pt
           \hbox{$>$}\hbox{$\sim$}}}}
\def\drawbox#1#2{\hrule height#2pt

\hbox{\vrule width#2pt height#1pt \kern#1pt
              \vrule width#2pt}
              \hrule height#2pt}

\def\Asym#1#2{\vcenter{\vbox{\drawbox{#1}{#2}
              \kern-#2pt       
              \drawbox{#1}{#2}}}}

\newcommand{\be}{\begin{equation}}
\newcommand{\ee}{\end{equation}}
\newcommand{\bea}{\begin{eqnarray}}
\newcommand{\eea}{\end{eqnarray}}

\newcommand{\neu}[1]{\ensuremath{\tilde{\chi}_{#1}^0}}

\newcommand{\gsim}{\lower.7ex\hbox{$\;\stackrel{\textstyle>}{\sim}\;$}}
\newcommand{\lsim}{\lower.7ex\hbox{$\;\stackrel{\textstyle<}{\sim}\;$}}

\newcommand{\ttbar}{t \bar{t}}

\newcommand{\met} {{E\!\!\!\!/_{\rm T}}}

\newcommand{ \pythia } {{\tt PYTHIA}}

\newcommand{ \madgraph } {{\tt MADGRAPH 5.11}}
\newcommand{ \delphes } {{\tt DELPHES 3.1.2}}
\newcommand{ \feynrules } {{\tt FEYNRULES}}
\newcommand{ \fastjet } {{\tt FASTJET}}
\newcommand{ \Prospino } {{\tt Prospino2.1}}

\def\bar {\overline}

\newcommand{\GeV}{\text{ GeV}}

\begin{document}

\title{Interpreting the CMS $\ell^+\ell^- jj \met$  Excess with a Leptoquark Model}

\author{Ben Allanach$^a$}

\author{Alexandre Alves$^b$}
\email{aalves@unifesp.br}

\author{Farinaldo S. Queiroz$^{c,d,e}$}

\author{Kuver Sinha$^f$}

\author{Alessandro Strumia$^{g,h}$}
\affiliation{$^a$DAMTP, CMS, Wilberforce Road, University of Cambridge, Cambridge, CB3 0WA, United Kingdom \\
$^b$Departamento de Ci\^encias Exatas e da Terra, Universidade Federal de S\~ao Paulo, Diadema-SP, 09972-270, Brazil \\
$^c$Department of Physics and Santa Cruz Institute for Particle
Physics University of California, Santa Cruz, CA 95064, USA\\
$^d$Max-Planck-Institut f\"ur Kernphysik, Saupfercheckweg 1, 69117 Heidelberg, Germany\\
$^e$International Institute of Physics, UFRN, Av. Odilon Gomes de Lima, 1722 - Capim Macio - 59078-400 - Natal-RN, Brazil
$^f$Department of Physics, Syracuse University, Syracuse, NY 13244, USA\\
$^g$Dipartimento di Fisica dell' Universita di Pisa and INFN, Italy\\
$^h$National Institute of Chemical Physics and Biophysics, Tallinn, Estonia}

\date{\today}

\begin{abstract}
We present a model of leptoquarks (LQs) with a  significant partial branching ratio into an extra sector, taken to be a viable dark matter candidate, other than the canonical lepton and jets final state. For LQs with mass around 500 GeV, the model reproduces the recent excess claimed by the CMS collaboration in $\ell^+\ell^- jj \met$ final state: the event rate, the distribution in di-lepton invariant mass and the rapidity
range are compatible with the data. The model is compatible with other collider bounds including LQ searches, as well as bounds from meson mixing and decays. Prospects of discovery at Run II of the LHC are discussed. 
\end{abstract}

\maketitle



\begin{figure}[!h]
\centering
\includegraphics[width=8.5cm]{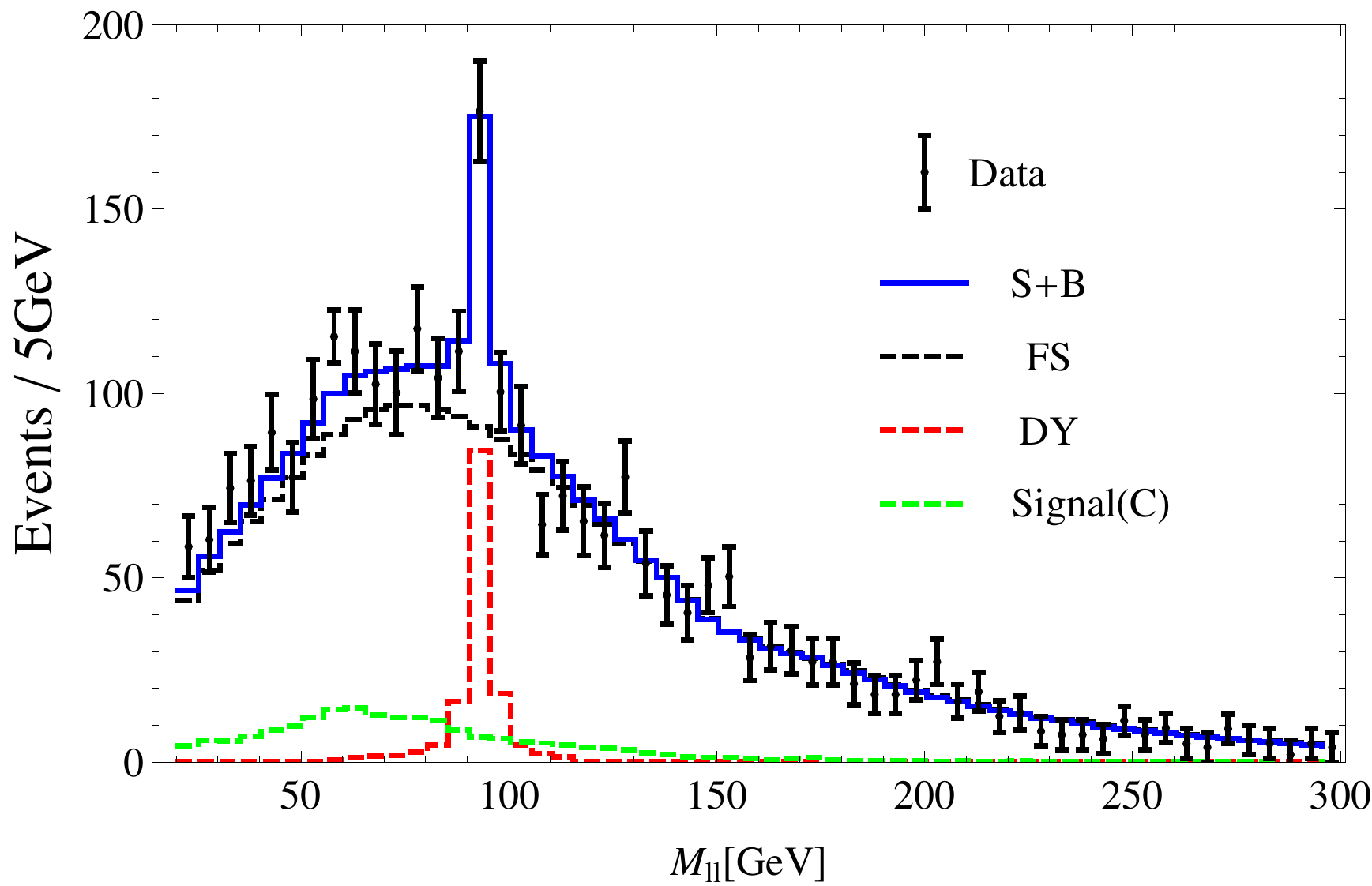}
\caption{Di-lepton invariant mass spectrum of CMS $\ell^+\ell^- jj \met$  events: expected non-DY background (black histogram), expected DY (red histogram), expected signal for benchmark C (green histogram) and expected signal plus background for benchmark C (blue histogram).$21$ signal events are expected for $m_{\ell\ell} > 100$ GeV, which is compatible with CMS data. }
\label{dileptondistri}
\end{figure}

{\bf Introduction.} The CMS collaboration reported a 2.6$\sigma$ excess compared with 
Standard Model expectations in 
$\ell^+\ell^- jj \met$ events, containing two 
opposite-sign same-flavor leptons $\ell^\pm =\{e^\pm,\mu^\pm\}$, at least two jets and missing transverse momentum $\met$~\cite{CMS-PAS-SUS-12-019} with 19.4 fb$^{-1}$ of integrated luminosity at a center of mass energy of 8 TeV.
The excess was found in the central region with lepton pseudo-rapidities
$|\eta_\ell | \, \leq \, 1.4$, after event selection and flavor
subtraction cuts, and 
with di-lepton invariant mass $m_{\ell\ell} <80\GeV$,
 as shown in Fig.~\ref{dileptondistri}. No excess is seen in other regions nor in the trilepton channel. 

The excesses were found in the context of searches for edges in $m_{\ell \ell}$.
The triangular edge is a classic supersymmetry signal. Interpretations of
the CMS excess in the context of the so-called golden cascade 
($\neu{2}\rightarrow \tilde{\ell}^{\pm} \ell^{\mp} \rightarrow \neu{1}\ell^{\pm}\ell^{\mp} $) 
have been proposed ($\neu{1}, \neu{2},$
and $\tilde{\ell}$ are the lightest neutralino, the next-to-lightest
neutralino and 
the slepton, respectively). Since direct electroweak production of $\neu{2}$
has a too small cross section
to provide a large enough rate whilst evading previous collider bounds from
LEP, assistance from colored particle production is 
required. The decay chain could start with $\tilde{t} \rightarrow t \neu{2}$
\cite{Dutta:2013sta} or $\tilde{q} \rightarrow q \neu{2}$
\cite{Allanach:2014gsa}. The former interpretation is constrained by the fact
that the CMS study did not observe a large excess in trilepton final states,
which should be present from the leptonic top quark decay. 
The CMS study itself opted for an
explanation in terms of light sbottoms $\tilde{b} \rightarrow b \neu{2}$, 
and $\neu{2}\to \neu{1}\ell^+\ell^-$, 
although no $b$-jet requirement was made on the final states. 
Ref.~\cite{Huang:2014oza} explored the parameter space in
order to simultaneously satisfy bounds from 4 charged lepton production. Recently, however, several potential SUSY scenarios explaining the CMS excess have been shown to be in tension with existing experimental data~\cite{Grothaus:2015yha}.

The purpose of this letter is to point out that the CMS excess can be
explained by a different, non-supersymmetric, class of models with leptoquarks (LQs) \cite{Pati:1974yy} - \cite{Gershtein:1999gp}.
%
In a model of LQs  presented in~\cite{Queiroz:2014pra} by some of the authors, the LQ had branchings to three possible final states: $(i)$ the usual final state with charged lepton and jet; $(ii)$ final state with jets and missing energy $\met$ in the form of dark matter (DM); and $(iii)$ final state with charged lepton, jets, and $\met$ in the form of DM. By a combination of branchings and kinematics, the model was able to account for the mild excess in  the $ee jj$ and $e \nu jj$ channels observed by a recent CMS search~\cite{CMSresult2,CMSresult} for first generation LQs in the mass range 550 - 650 GeV. To account for the event rate observed by CMS in these first generation LQ searches, the branching ratio of the LQ into electrons and quarks was taken to be  $\approx 15\%$. A viable DM sector was constructed which accounted for the remaining branching ratio through channels $(ii)$ and $(iii)$ mentioned above.





In this work, we note that the same framework, through its connection to DM, interestingly enables us to fit the CMS $\ell^+\ell^- jj \met$ excess. 
Fig.~\ref{dileptondistri} exemplifies how the $\ell^+\ell^- jj \met$ excess can be reproduced
by LQs with $\approx 500$ GeV masses:
the model predicts a peak in $m_{\ell \ell}$, which fits data roughly as
well as  the triangular edge searched for by CMS.   
The model also predicts small numbers of events in the forward region
and for trilepton final states, also compatible with the CMS measurements. The model is compatible with flavor constraints, as well as constraints from other LHC searches.
We will see that the benchmark point that best fits the CMS $\ell^+\ell^- jj \met$ excess is around $\sim 500$ GeV where there is no excess in the $ee jj$ and $e \nu jj$ channels. This necessitates a different choice of branchings and benchmark from ~\cite{Queiroz:2014pra}.

{\bf The Model}. \label{model} 
We consider LQs that do not lead to proton decay at renormalizable level:
either a scalar $R_2$ in the $\left(3,2, 7/6\right)$ representation of ${\rm SU}(3) \otimes {\rm SU}(2) \otimes {\rm U}(1)$
that can couple to $\bar{Q}e$ and to $ L \bar{u}$;
or a scalar $\tilde R_2$ in the $(3,2, 1/6)$ representation (the same as the quark doublet $Q$)
that can couple to $L\bar d$ \cite{Buchmuller:1986zs,Davies:1990sc}. We note that dimension five operators involving these LQs can lead to proton decay; these have to be forbidden by a discrete symmetry, as shown in detail in ~\cite{Queiroz:2014pra}.

We focus on $\widetilde{R}_2$ because its quantum numbers are favorable from a DM perspective, as we will clarify later. Pair production of $\tilde R_2$ via gluon fusion and quark-antiquark annihilation thus constitutes our main example.
Its Lagrangian couplings are
\bea \label{lepto_lag1}
 -\lambda_d^{ij} \bar{d}_{R}^i \widetilde{R}_2^T\epsilon L_L^j + {\rm h.c.}\ ,
\eea
where $i,j$ denote flavor indices. In the above,
\bea
\widetilde{R}_2 = \left(
      \begin{array}{c}
        V_\alpha \\
        Y_\alpha \\
      \end{array}
    \right) \ , \ \ \ \epsilon = \left(
                                   \begin{array}{cc}
                                     0 & 1 \\
                                     -1 & 0 \\
                                   \end{array}
                                 \right) \ ,  \ \ \ L_L = \left(
                                                            \begin{array}{c}
                                                              \nu_L \\
                                                              \ell_L \\
                                                            \end{array}
                                                          \right)\ .
\eea

Expanding the ${\rm SU}(2)$ components yields
\bea
-\lambda_d^{ij} \bar{d}_{\alpha R}^i(V_\alpha e_L^j - Y_\alpha \nu_L^j)  + {\rm h.c.}
\eea

In addition to the LQ, we introduce a dark sector with a scalar $S$ and a fermion $\chi$, with a Z$_2$ symmetry under which the dark sector is odd, whereas the SM and LQ sector is even. Thus, our new physics content is
\bea
\widetilde{R}_2 &=& (3,2,1/6)_{+} \nonumber \\
S  &=&  (1,3,0)_{-}  = \ \left( \begin{array}{cc}
\frac{1}{\sqrt{2}} S^0 & S^{+} \\
S^{-} & - \frac{1}{\sqrt{2}} S^0
 \end{array} \right) \, ,\\
\chi &=& (1,1,0)_{-} \,\,,
\eea
where we have also displayed the Z$_2$ quantum numbers as a subscript. 
Notice that there is almost no freedom in choosing the above particle content. The quantum charges of the dark sector are fixed by symmetries; either the scalar $S$ or the fermion $\chi$ has to be a triplet under SU$(2)_L$ (having both singlets will not give enough charged leptons in the final state to match the CMS study, as we will see). We have chosen a scalar triplet for convenience; the discussion and results would be analogous for a model with a triplet fermion $\chi$ and a singlet scalar $S$. The hypercharge of the dark sector is fixed to be zero to easily accommodate DM direct detection limits. 

\medskip

The LQ decay into the dark sector can then be described by adding the following dimension-5 effective operators:
\be \label{modelL1}
- \, \frac{h_i}{\Lambda_1}  S \overline{Q}_i \chi \widetilde{R}_2 \,- \, \frac{h^{\prime}_i}{\Lambda_2}  S \overline{\ell}_i \chi \tilde{H}\,\, ,
\ee
%
 where $\tilde{H}$ is the iso-spin transformation  of the Higgs doublet~\cite{Queiroz:2014pra}.

The CMS study does not discuss the relative number of $e^+e^-$ and $\mu^+ \mu^-$ events in the $\ell^+\ell^- jj \met$ excess. We will assume them to be equal and introduce both a first generation LQ and a second generation LQ in order to obtain both $e^+e^-$ and $\mu^+ \mu^-$ events. In the following, however, we will describe our methods for first generation LQs, keeping in mind that second generation LQs can be similarly treated by a simple replacement of electrons by muons.

{\bf Spectrum and Decays.} Either the triplet component $S^0$ or the singlet $\chi$ can be the lightest
DM particle.  The two possibilities lead to distinct DM phenomenology. Here, we will mainly consider the case of a singlet $\chi$ as the DM candidate.

The two couplings of the LQ induce two decay modes:$\widetilde{R}_2 \, \rightarrow \, e j\qquad  \hbox{and} \qquad \widetilde{R}_2 \rightarrow Sj \chi$. The latter coupling  in Eq.~(\ref{modelL1}) induces the $S\to e \chi$ and $S\to \nu \chi$ decays of $S$.  In components: $S^{\pm} \, \rightarrow \, \chi e^\pm \,\,\,, \qquad S^0 \to \nu \chi$, so that decays of charged scalars $S^{\pm}$ give charged leptons and $\met$ in the final state. One loop electro-weak corrections induce a small mass splitting of $\approx 200$ MeV between the neutral and charged states in $S$. 
%
%
Combining all decays, one $\tilde{R}_2$ LQ can produce the following final states:
\begin{enumerate}
\item A charged lepton and a jet.
The free couplings of the Lagrangian 
allow us to set the $\widetilde{R}_2$ branching ratio to the level required to be compatible with CMS searches for first and second generation LQs, with final states $ee jj$ and $e \nu jj$, and similarly for muons. We will find that at the benchmark point of this study, there is no excess in these channels; thus, we will choose
$${\rm BR}(\widetilde{R}_2 \, \rightarrow \, e j) \, \sim \,  0\% \, ,$$ 
consequently ${\rm BR}(\widetilde{R}_2 \rightarrow S j \chi)\, \sim \, 100\% $. 

\item A jet and missing energy ($\met $), with 
$${\rm BR}(\widetilde{R}_2  \to S^{0} j \chi \to j\met) \sim 29\% .$$

\item A charged lepton, a jet and missing energy, with 
$${\rm BR}(\widetilde{R}_2  \to S^{\pm} j \chi \to  \ell^\pm j \met)\sim 71\% .$$
\end{enumerate}



{\bf Constraints on the masses of $\widetilde{R}_2$, $S$, $\chi$.} $(i)$ Constraints from $j  \met$ searches. 
LQ pairs decaying to jets and missing energy must be compatible with $j  \met$ searches from ATLAS \cite{Aad:2014wea}
and CMS \cite{Chatrchyan:2014lfa}. The most relevant exclusion limits from
these searches are phrased in the squark-neutralino $(m_{\tilde{q}},
m_{\neu{1}})$ mass plane (Fig.~10 of \cite{Aad:2014wea}), where the lightest
neutralino is stable and escapes the detector. In our  model, both $S^0$ and
$\chi$ escape undetected. To compare to the experimental studies,
we map the neutralino mass $m_{\neu{1}}$ to  $m_S + m_{\chi}$ and the squark
mass $m_{\tilde{q}}$ to $m_{\widetilde{R}_2}$.  
Taking into account that, in our case, ${\rm BR}(\widetilde{R}_2 \rightarrow j
\met) \sim 29\%$, 
we conservatively estimate the bound
%
\be
m_S + m_{\chi} \, > \, 300 \, {\rm GeV} 
\ee
for LQs in the mass range $450-650$ GeV. 

%
%
%
%

\bigskip

$(ii)$ Constraints from CMS LQ searches: There is a mild  evidence for  $\sim \, 550 - 650$~GeV first generation scalar LQs, using $19.6 {\ \rm fb^{-1}}$ of integrated luminosity, at $2.4 \sigma$  and $2.6 \sigma$, in the $eejj$ and $e \nu jj$ channels respectively \cite{CMSresult}.  A branching ${\rm BR}(\widetilde{R}_2 \, \rightarrow \, e j) \, \sim \, 15\%$ agree well with the observed excess in the $550-650$ GeV mass range. For second generation LQs, studies of the $\mu \mu jj$ and $\mu \nu jj$ final states has resulted in the exclusion of scalar LQs with masses below 1070 (785) GeV for ${\rm BR}_{\mu j} = 1 (0.5)$~\cite{CMSresult2}.



As stated before, we will assume a branching  ${\rm BR}(\widetilde{R}_2 \, \rightarrow \, e j) \, \sim \, 0\%$ to switch off decays to this channel for LQs of mass around 500 GeV.  We will also assume ${\rm BR}(\widetilde{R}_2 \, \rightarrow \, \mu j) \, \sim \, 0\%$ to be compatible with second generation LQ searches.

%
%
%


\bigskip

$(iii)$ Constraints from the dilepton invariant mass $m_{\ell\ell}$ distribution of the CMS $\ell^+\ell^- jj \met$ excess,
located mostly below  $m_{\ell\ell}\approx 80$~GeV. 
The two leptons in our case come from the decay $S \rightarrow \ell \chi$. To get the excess in the required range  the mass difference between $S$ and $\chi$ should be $m_S - m_{\chi} \, \sim \, 20-40 \, {\rm GeV} \,\,.$
Spectra where $m_S - m_\chi \gsim 40$ GeV are disfavored since the dilepton
invariant mass distributions would peak 
at a value of $m_{\ell\ell}$ that is too large compared to the excess.
On the other hand, for $m_S - m_\chi \lsim 20$ GeV, the leptons are too soft
and do not survive the $p_T$ cuts for leptons. \\

{\bf Results.} \label{results} We take our background estimates from \cite{CMS-PAS-SUS-12-019}. Opposite sign opposite-flavor (OSOF) leptons from $\ttbar$, which has the same rate of the same-flavor (OSSF) channel, are used to measure these backgrounds in the CMS study. Drell-Yan production, the main irreducible background, is estimated by a control region which does not overlap with the signal region. 

We follow the CMS counting experiment analysis in \cite{CMS-PAS-SUS-12-019}
for the signal. The final state is required to have at least two leptons and
at least two jets. The cuts employed are: 

Cut $(i)$ Two OSSF leptons are required to be present, with $p_T > 20$ GeV in $|\eta|<1.4$ which is defined as the {\it central region} in the CMS study. 

Cut $(ii)$ At least two jets are required with  $p_T>40$ GeV in
$|\eta|<3.0$. 

We use \fastjet~\cite{Cacciari:2011ma} to reconstruct jets. An event is selected if it contains two
jets and satisfies $\met > 150$ GeV, or, if it contains three or more jets and
satisfies $\met > 100$ GeV. In the dilepton invariant mass range  20 GeV $<
m_{\ell \ell}<70$ GeV, the total background estimate provided by the CMS study
for central OSSF events is 730$\pm 40$. The observed number 
was 860, corresponding to an excess of $130^{+48}_{-49}$ events provided
by new physics, we hypothesize. In our model, this excess number of events is
produced by first and second generation LQs. 
We implement the model in \feynrules ~\cite{Christensen:2008py} and calculate
the branching ratios and cross sections using \madgraph
~\cite{Alwall:2011uj}. The events are then passed onto \pythia
~\cite{Sjostrand:2006za} for parton showering and hadronization followed by
the modelling of detector effects by  \delphes ~ \cite{deFavereau:2013fsa}. We
took exactly the same electron/muon/jets isolation criteria adopted
in~\cite{CMS-PAS-SUS-12-019}. Production cross sections were normalized by the NLO QCD rate using \Prospino~\cite{Prospino2.1}.

We performed a scan over the masses $(m_{\widetilde{R}_2}, \,\, m_{S}, \,\, m_{\chi})$, fixing 
the Lagrangian parameters $\lambda^{ij}_d= 10^{-4}$, $h_i/\Lambda_1 = 10^{-3}$  GeV$^{-1}$ and $h_i^\prime/\Lambda_2= 10^{-3}$ GeV$^{-1}$. In the above, we have assumed democratic values for the Yukawas corresponding to $i = 1,2$ and $j=1,2$. We note that constraints arising from meson decays and mixings are sensitive to products of the Yukawas $\lambda^{ij}$; in particular, our choice of Yukawas is compatible with bounds given in \cite{Dorsner:2014axa} and  \cite{Queiroz:2014pra}. The cut flow is displayed in Table~\ref{cutflow} for some spectra that best
fit the number of events in the central region of the CMS search with ${\rm
  BR}(\widetilde{R}_2 \, \rightarrow \, ej) \, \sim \, 0\%$, $m_S+m_\chi >
300$ GeV and $m_S-m_\chi<40$ GeV. 

We have assumed theoretical uncertainties following \cite{Kramer:2004df}, which are the same as the estimates used by CMS in its LQ search \cite{CMS-PAS-SUS-12-019}. The uncertainties are given in Table 1 of the CMS paper \cite{CMS-PAS-SUS-12-019}. Variation of the renormalization/factorization scale between half and twice the LQ mass leads to a $\sim 25\%$ uncertainty in the production cross section. The PDF uncertainty is approximately $20\%$ of the NLO cross section. This can easily lead to a $\sim 20\%$ variation in the number of signal events, which will be interesting to pursue if the signal becomes stronger in the next run.

\begin{table}[!htp] 
\caption{Summary of the effective cross-sections (fb) for some benchmark signal points that best fit the CMS signal at LHC8. In the fourth and fifth rows of each point we also show in parenthesis the final number of events predicted.  The last row displays the number of events in the forward region. Masses are in GeV. Lagrangian parameters are fixed at $\lambda^{ij}_d= 10^{-4}$, $h_i/\Lambda_1 = 10^{-3}$  GeV$^{-1}$ and $h_i^\prime/\Lambda_2= 10^{-3}$ GeV$^{-1}$. The benchmark point for the current study is $C$. 
The cross sections (in fb) predicted at LHC14 after imposing the cuts of the LHC8 analysis for the three best benchmark points is displayed in the last column.}
\label{cutflow}
\begin{center}
\begin{tabular}{c c c c} 
\hline\hline 
$(m_{\widetilde{R}_2}, m_{S}, m_{\chi})$ & Selection & Signal (fb) & Signal$_{14}$ (fb)\\
\hline\hline \\

               &  preselection  & 149.1      \\
         $A:\; (450,200,170)$  &   Cut $(i)$   & 40.2  & 46.6 \\
           &     Cut $(ii)$ & 10.5  \\
                &      $20 < m_{\ell \ell} < 70$ & 5.7(110) \\
           &      $m_{\ell \ell} > 100$ & 1.8(37) \\
           &   Forward & 18 \\
                         \hline \\

                        \hline \\
              &  preselection  & 105.7      \\
         $B:\; (500,200,170)$  &   Cut $(i)$   & 28.5   & 69.6 \\
           &     Cut $(ii)$ &  11.6 \\
                &      $20 < m_{\ell \ell} < 70$ & 5.7(112) \\
            &      $m_{\ell \ell} > 100$ & 2.3(45) \\
           &   Forward & 14 \\
                        \hline \\

                   &  preselection  & 126.1     \\
         $C:\; (500,160,140)$  &   Cut $(i)$   & 19.0  & 53.6 \\
           &     Cut $(ii)$ &  10.0 \\
                &      $20 < m_{\ell \ell} < 70$ & 5.8(114) \\
            &      $m_{\ell \ell} > 100$ & 1.11(21) \\
            &   Forward & 8 \\
                         \hline \\

                         \hline \hline \\

\end{tabular}
\end{center}
\end{table}

%
%


In  Fig.~\ref{dileptondistri}, we show the comparison between the CMS data points and the predicted distribution for our model at benchmark point $C$. As we see, our model can fit the data very well. In fact, points A,B and C of Table~\ref{cutflow} present this feature. While the CMS study showed a triangular shape with a sharp edge, for benchmark point $C$ around $21$ events survive in the region $m_{\ell \ell} > 100$ GeV, which is well within the background uncertainty.  
%

In order to better quantify the agreement between these models predicitions and the data we used the $\chi^2$ statistics defined below in Eq.~\ref{chi2} which, as in the experimental study, compares a model prediction in the Central ($|\eta_\ell|<1.4$) and Forward ($1.6<|\eta_\ell|<2.4$) regions
\begin{eqnarray}
\chi^2(\mu)&=&\min_{\{\mathbf{\theta}\}} \sum_{i=1}^{N_{bin}}\left[\frac{(\mu s^I_i+\mathbf{\theta}\cdot \mathbf{b^I}_i-d^I_i)^2}{(\sigma^I_i)^2}\right.\nonumber \\
&+& \left.\frac{(\mu s^O_i+\mathbf{\theta}\cdot \mathbf{b^O}_i-d^O_i)^2}{(\sigma^O_i)^2}\right]\nonumber \\
&&
\label{chi2}
\end{eqnarray}
In this formula, $s^{I(O)}_i$, $\mathbf{b^{I(O)}}_i$ and $d^{I(O)}_i$ denote the $i$-th dilepton invariant mass bin for the signal, backgrounds(DY and FS) and data distributions, respectively, in the Central region. The superscipts $I$ and $O$ denotes the Inside the window region $20<m_{\ell\ell}<70$ GeV and the Outside region $m_{\ell\ell}>70$ GeV, respectively. 
The corresponding experimental uncertainties in each bin are given by $\sigma_i^I$, $\sigma_i^O$ and $\sigma^F$. In order to better approach the uncertainties in the background bins we take into account two nuisance parameters $\mathbf{\theta}=(\theta_1,\theta_2)$ for the DY background normalization, $b_1$, and the FS background normalization $b_2$. 

 All these regions are crucial to determine the pattern of decays and the spectra favored by the data. For example, those spectra where the $m_S-m_\chi$ are bigger than $\sim 40$ GeV are disfavored in this respect as their dilepton invariant mass distributions peak outside the window. On the other hand, for $m_S-m_\chi\lesssim 20$ GeV, the leptons are too soft and do not survive the $p_T$ cuts for leptons. We show in Fig.~\ref{mllC} the dilepton invariant mass distribution for the point $C$. The CMS study compared their data against a $R-parity$ conserving supersymmetric scenario where an sbottom decays to a SF lepton pair and missing energy. This type of decay presents a dilepton mass distribution with a triangular shape and a kinematic edge around $m_{\tilde{b}}-m_{\chi_2^0}$ which fits the observed excesses well. 

The Forward region is an important control region since it is expected that heavy particles undergoing decays as in SUSY or LQ models (as suggested in this work) will produce central leptons. We checked that the number of events is small for the best LQ models, around 7 events for benchmark $C$ and similarly for the others as shown in Table~\ref{cutflow}. This is consistent with the CMS reported number of $6 \pm 20$ events in the forward region. Moreover, the
expected fraction of signal events with 3 or more leptons in those benchmark
points is never beyond $0.5\%$, again consistent with the absence of an excess in the
trilepton channel. 

We show in Table~\ref{xi2} the $\chi^2$ obtained with Eq.~(\ref{chi2}) and the corresponding $p$-value calculated as $p=\int_{\chi^2_C}^{\infty}\chi^2_{n-1}(x)dx$, where $n=60$ degrees of freedom, for the LQ models, the SUSY model of the experimental study and the pure background model. The numbers in parenthesis represent the one-sided probability in the tail of a Normal distribution. The CMS study assigns a 2.4$\sigma$ significance for the SUSY model against 2.6$\sigma$ from our estimate. The discrepancy might be due the estimation of the errors included in the fit which are not fully accessible to us. Nevertheless, we also found that the best fit is provided by a SUSY type signal and point B. The benchmark points A, B and C, by their turn, are the best fitting points in our model as can been seen in Table~\ref{xi2}, and very similar to the SUSY fit.
%
\begin{table}[h!]
\begin{center}
    \begin{tabular}{ c c c }
    \hline \hline
     Model &  ~~$\chi^2$~~ &  $p$-value \\ 
     \hline
    A  & 35.0 & 0.005(2.5$\sigma$) \\ 
    B  & 34.3 & 0.004(2.6$\sigma$) \\ 
    C  & 35.0 & 0.005(2.5$\sigma$) \\ 
     SUSY & 34.3 & 0.004(2.6$\sigma$) \\
     Background & 40.6 & 0.03(1.8$\sigma$) \\
\hline\hline \\
    \end{tabular}
\end{center}
\caption{\footnotesize{The $\chi^2$ statistics computed from Eq.~(\ref{chi2}) for three benchmark LQ models, the SUSY model and the background model taken from the CMS analysis~\cite{CMS-PAS-SUS-12-019}. In the third column we show the $p$-value of each model and the corresponding statistical significance in parenthesis as described in the text.}}
\label{xi2}
\end{table}

With the current data, none of these models can be excluded at 90\% CL, for example, although some models can provide a better fit than others. The background model is currently the worst fit amongst all of them. Comparing the LQ and supersymmetry interpretations we see that the SUSY model is able to explain the data as well as our best point, B. The benchmark point C, however, is the one that better fits the excess quoted in the CMS work in the signal region $20<m_{\ell\ell}<70$ GeV, 114 events against 130 of the CMS fit. Moreover, it predicts the smallest number of events in the Forward region of all benchmark points, 8 events against 6 from CMS.



\begin{figure}[!h]
\centering
\includegraphics[width=8.8cm]{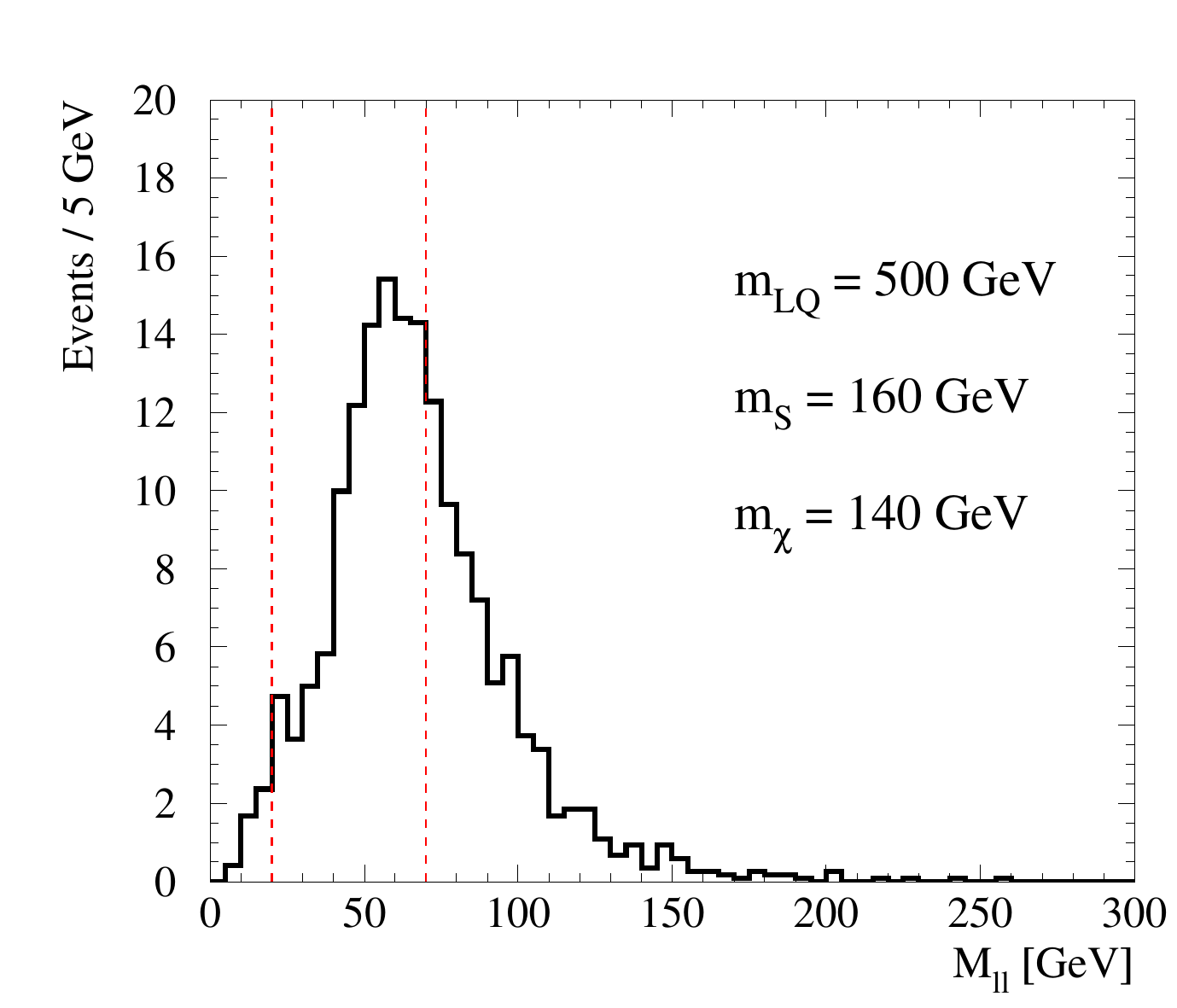}
\caption{The dilepton invariant mass distribution of the signal events (point $C$) is displayed. The vertical dashed lines indicate the invariant mass cut. Around $21$ events survive in the region $m_{\ell \ell} > 100$ GeV, which is well within the background uncertainty.}
\label{mllb}
\end{figure}
%


In the last column of Table~\ref{cutflow} we show the projected cross sections at the LHC14 surviving the cuts of the LHC8  analysis for the three best benchmark points found.\\
%


%

{\bf Comments on Dark Matter Phenomenology}. DM stability  is guaranteed by the discrete Z$_2$ symmetry. The options are to have either triplet or singlet DM that could be the fermion $\chi$ or the scalar $S$ as discussed thoroughly in \cite{Queiroz:2014pra}. Taking into account, direct, indirect and collider constraints the following possibilities emerge:
\begin{enumerate}
\item A scalar singlet $S$ coupled to the Higgs portal. After taking LUX, indirect detection and Higgs invisible decay widths into account the mass range $m_S \, > \, 100$ GeV and $m_S \, \sim \, 60-65$ GeV is allowed. However, XENON1T is expected to masses up to a 1 TeV \cite{Queiroz:2014yna}.
\item A fermionic singlet ($\chi$) with a significant pseudoscalar coupling to the Higgs portal results into a less constrained spin-dependent scattering cross section \cite{Fedderke:2014wda}. With $\Lambda$ being the scale of the portal interaction, and $\sin \xi$ the pseudoscalar coupling, from Fig.7 of \cite{Fedderke:2014wda}, it is clear that $m_\chi \, \gsim \, \mathcal{O}(100)$ GeV is allowed by data for $\sin^2 \xi \, \gsim \,0.7 $ with $\Lambda \sim 1-5$ TeV. The bounds do not depend much on whether $\chi$ is Majorana or Dirac.

\end{enumerate}

Thus, a singlet fermionic DM candidate with mass $m_\chi \sim 140$
GeV that we have taken in our collider analysis of this paper is currently a
viable option. \\

{\bf Conclusions.} In this paper, we have shown that the excess observed by CMS in the $\ell^+\ell^- jj \met$ search can be explained consistently within a class of LQ models.
%
We particularly discussed a model which consists of scalar first generation and second generation LQs that decay dominantly to leptons, jets, and missing energy in the form of a stable DM candidate. The confluence of proton decay constraints and DM direct/indirect detection results in a highly predictive model of LQs that can satisfy the CMS  $\ell^+\ell^- jj \met$ search. Our model provides a general proof of concept that a peak distribution in the required dilepton mass range can be obtained outside the purview of supersymmetry.

Benchmark point C consists of first and second generation LQs with masses of
$500$ GeV, and dark sector particle masses of $m_S = 160$ GeV and $m_{\chi} =
140$ GeV. The LQs dominantly decay to $\ell j \met$ final states ($\sim 71\%$
branching), and subdominantly into $j \met$ final states ($\sim 29\%$ branching), while there is negligible branching into the canonical LQ final states of $\ell j$.


While CMS
fit a triangular shape in the opposite-sign-same-flavor dilepton invariant
mass distribution after event selection and flavor subtraction, it is too
premature to settle definitively on a kinematic edge. In our model, the number
of signal events in the window $20 \, < m_{\ell\ell} \, < \, 70$ GeV after
event selection is $114$ for the benchmark point. The dilepton mass distribution peaks in the window
between $20 - 70$ GeV with the required event count, while the number of
events in the region $m_{\ell\ell} \, > \, 100$ GeV is within the background
uncertainty. In a  simple $\chi^2$ fit, this point compares favorably with the CMS SUSY fit in \cite{CMS-PAS-SUS-12-019}. 
 The model is consistent with the non-observation of signal in the
forward region, and in the trilepton final states. We have also provided projections for the signal cross sections
at LHC14 in the best benchmark models. For the benchmark point $C$, for example, $\sim 5360$ events are expected with $100$
fb$^{-1}$ luminosity.


These results highlight the importance of refining the search strategy and that a potential LQ
discovery is attainable in the next LHC run. \\

\section*{Acknowledgements}

This work was supported by Fundac\~ao de Amparo \`a Pesquisa do Estado de
S\~ao Paulo (FAPESP) grant 2013/22079-8, STFC grant ST/L000385/1, 
US Department of Energy Award SC0010107 and the Brazilian National Counsel for
Technological and Scientific Development (CNPq) grant 307098/2014-1(AA), and
NASA Astrophysics Theory Grant NNH12ZDA001N, by ESF grant MTT8.


\end{document}